# SPECTRAL CLASSIFICATION OF GALAXIES ALONG THE HUBBLE SEQUENCE

Dennis Zaritsky

UCO/Lick Observatory and Board of Astronomy and Astrophysics, Univ. of California, Santa Cruz, CA, 95064, E-mail: dennis@lick.ucsc.edu

Ann I. Zabludoff and Jeffrey A. Willick

Carnegie Observatories, 813 Santa Barbara St., Pasadena, CA, 91101, E-mail: aiz@ociw.edu and jeffw@ociw.edu

## ABSTRACT

We develop a straightforward and quantitative two-step method for spectroscopically classifying galaxies from the low signal-to-noise (S/N) optical spectra typical of galaxy redshift surveys. First, using $\chi^2$-fitting of characteristic templates to the object spectrum, we determine the relative contributions of the old stellar component, the young stellar component, and various emission line spectra. Then, we classify the galaxy by comparing the relative strengths of the components with those of galaxies of known morphological type. In particular, we use the ratios of (1) the emission line to absorption line contribution, (2) the young to old stellar contribution, and (3) the oxygen to hydrogen emission line contribution. We calibrate and test the method using published morphological types for 32 galaxies from the long-slit spectroscopic survey of Kennicutt (1992) and for 304 galaxies from a fiber spectroscopic survey of nearby galaxy clusters. From an analysis of a sample of long-slit spectra of spiral galaxies in two galaxy clusters, we conclude that the majority of the galaxies observed in the fiber survey are sufficiently distant that their spectral classification is unaffected by aperture bias. Our spectral classification is consistent with the morphological classification to within one type (e.g. E to S0 or Sa to Sb) for $\gtrsim 80\%$ of the galaxies. Disagreements between the spectral and morphological classifications of the remaining galaxies reflect a divergence in the correspondence between spectral and morphological types, rather than a problem with the data or method.

## 1. INTRODUCTION

Classification is a basic step toward understanding nature. The classification of galaxies provides a framework within which to understand how galaxies form and evolve. The fundamentals of the morphological classification scheme for galaxies as proposed by Hubble (1926, 1936) are still in place, despite some modifications and additions to the basic sequence (cf. de Vaucouleurs 1959; Sandage 1961). The intrinsic value of the morphological sequence arises from the strong correlations between morphological type and many physical characteristics of galaxies (cf. Roberts & Haynes 1994). Spectral classification, pioneered by Morgan and Mayall (1957), provides an alternative to classification based on images. The principal advantage of this approach is that it is based directly on the physics of the stellar populations and the interstellar medium.

The large volume of data and the need for uniform, reproducible, and physically-oriented galaxy classification motivate the search for an efficient spectral classification method. Such a method, based on the stellar populations and emission-line properties of galaxies, must be successful even when applied to the low S/N spectral data typical of current galaxy redshift surveys. Spectral classifications correlate well with morphological classifications (Morgan & Mayall 1957), but also complement them by providing information not available solely from the morphology.

We revisit spectral classification armed with digital data, computers, and potentially thousands of galaxy spectra. Parallel efforts are underway (*e.g.*, Heyl 1994; Connolly *et al.* 1995; Bershady 1995). The procedure described by Heyl is potentially quite similar to ours, although his classification is based currently only on the galaxies in the Kennicutt (1992) sample, which, because of its small size, may not describe the full range of galaxy properties within any given galaxy type. Additionally, our technique has some technical advantages, such as the independent determination of emission and absorption line strengths. The technique developed by Connolly *et al.* primarily utilizes the shape of the continuum (and its sensitivity to recent star formation) to classify galaxies. As is evident from their Figure 3, the major difference between the two principal "eigen-spectra" used to classify their galaxies is the continuum behavior blueward of 4000 Å. This difference arises from the combination of the 4000 Å break and UV continuum emission from young stars. Their technique is optimized for flux-calibrated spectra that extend over a large wavelength region and is best suited to spectra that extend blueward of 4000 Å in the rest frame, such as those of high-redshift galaxies, but generally applicable to any set of galaxy spectra. Bershady's method is based on galaxy colors, which are observationally more efficient for samples of galaxies that are concentrated on the sky (such as high redshift or cluster galaxies), and bypasses the problem of aperture bias discussed below. On the other hand, some degeneracy exists in the colors between the line and continuum contributions, the colors need to be K-corrected, and photometric observations are needed in a variety of colors.

In this paper, we describe our method for spectroscopically classifying nearby galaxies and examine the correspondence between spectral and morphological classification. The principal difference between previous work and this study is our emphasis on the spectral lines rather than on the continuum shape (although we use some "continuum features", such as the 4000 Å break which is actually dominated by Ca II H & K absorption). Our philosophy is to keep the approach as simple as possible (adopting the fewest fitting parameters and the smallest number of classification criteria necessary to account for the dominant spectral signatures of different morphological types), so that the method is stable and can be applied to low S/N data. We fit characteristic template spectra to the object spectrum in order to determine the contributions of different spectral components. We then classify each galaxy using three ratios formed from the relative contributions of an early-type star (A), a late-type star (K), and an emission-line component. Our method can be applied uniformly and reproducibly to large numbers of the low S/N spectra typical of redshift surveys, which creates an opportunity to examine the properties of galaxies as a function of environment on an unprecedented scale. One potential limitation of the method, which we refer to as aperture bias, is that the small apertures characteristic of fiber or multislit observations may incompletely sample the galaxy. Therefore, we investigate the effect of aperture bias on our spectral classifications. We conclude that the majority are unaffected by aperture bias.

In §2, we describe the three spectroscopic samples used to calibrate and test the method. In §3, we discuss the classification algorithm, the application to the spectra, and the correspondence between spectral and morphological classification. In §4, we summarize our conclusions. Two appendices contain detailed discussions of the S/N limitations and the effect of aperture bias.

## 2. THE DATA

In this section, we describe the three different samples of galaxies used to calibrate and test our method of spectral classification. The first is a compilation of nearby and well-studied galaxies from a spectroscopic survey by Kennicutt (1992). We refer to this sample as the Nearby Galaxy (NG) sample. These are high S/N spectra obtained by allowing the galaxy to drift across the spectrograph slit. This observational technique produces spectra that are luminosity-weighted averages and representative of the entire galaxy. Therefore, although these galaxies are typically very nearby, the NG sample does not suffer from aperture bias. The sample's shortcoming is that its small size prevents us from fully sampling the variations in the correspondence between spectral and morphological type for any one type of galaxy.

The 55 original spectra in the NG sample, kindly provided by R. Kennicutt, were obtained mostly at the Steward 2.3m telescope and cover the wavelength region from 3650Å to 7100Å. The data are flux calibrated, but are not corrected for either Galactic or internal reddening (for details see Kennicutt 1992). The spectra include all of the major optical lines of galaxy spectra, except the Ca triplet at about 8500Å (cf. Figure 1 for example spectra). The spectral resolution is between 5 and 8 Å, which is comparable to that typically obtained in redshift surveys.

For the purpose of calibrating and testing our classification algorithm against the NG sample, we study only the 32 "normal" galaxies, where "normal" denotes a lack of broad emission lines. These are the first 27 galaxies in the Kennicutt catalog (1992; see his Table 1), plus NGC 2798, NGC 3034, NGC 3077, NGC 3310, and NGC 5195. Morphological types for these galaxies are drawn from Kennicutt's Table 1 and are from either the Revised Shapley-Ames Catalog (Sandage & Tammann 1981), the Second Reference Catalog of Bright Galaxies (de Vaucouleurs, de Vaucouleurs, & Corwin 1976), Huchra (1977), or Kennicutt (1992). Our rebinning of these morphological types into a coarser grid is outlined in Table 1. Based on the published morphological classifications and our rebinning of types, the NG sample contains 4 E's, 3 S0's, 5 Sa's, 5 Sb's, 9 Sc's, and 6 Irr's.

Table 1 : Our Rebinning of Galaxy Types

| Our Type | Standard Type | T-Type |
|---|---|---|
| E | E(dwarfs and giants) | -5,-4 |
| S0 | L, S0 | $-3,-2,-1$ |
| Sa | S0/a,Sa | 0,1 |
| Sb | Sab,Sb | 2,3 |
| Sc | Sbc,Sc | 4,5 |
| Irr | Scd,Sd,Sm,Irr | 6−10 |

Our second and much larger sample consists of 304 new fiber spectra of galaxies with published morphological classifications from Dressler (1980) and Richter, Materne, & Huchtmeier (1987) in the fields of nearby galaxy clusters. We refer to this sample as the Cluster (CL) sample. This sample is a factor of 10 larger than the NG sample and thus allows us to statistically examine the correspondence between spectral and morphological classification for galaxies of each type. This sample's potential shortcomings are that classification errors could result from either the limited fiber aperture size (aperture bias) or the low S/N of the spectra and that the spectral coverage (3300Å to 6200 Å) does not extend redward to H$\alpha$. However, because these data are very similar in quality to those of redshift surveys of large-scale structure, this sample provides a fair test of our spectral classification algorithm.

The CL sample spectra, which are part of a larger survey of the internal dynamics of rich clusters of galaxies (Zabludoff 1995), were obtained at the Las Campanas Observatory (LCO) with the 2.5m telescope and multifiber spectrograph (Shectman et al. 1992). The recessional velocities of galaxies in the survey are mostly between 3000 and 50000 km s$^{-1}$. The fiber aperture of 3 arcsec corresponds to a range of physical sizes between $\sim 0.6$ and 10 kpc, respectively ($H_0$ = 75 km s$^{-1}$ Mpc$^{-1}$). The galaxies have B-band apparent magnitudes between $\sim$ 15 and 18.5, and the total exposure times are between 2 and 6 hours. These data are not flux calibrated, which is typical of the redshift survey data to which our classification method might be applied. The lack of flux calibration prevents us from using the continuum shape as a classification criterion.

Figure 1: Characteristic spectra from the Kennicutt (1992) sample of normal galaxies of types E to Irr. The wavelength coverage of each panel is from 3600 to 7100Å. Even among adjacent types there are noticeable spectral differences.

The third sample is a random subsample of 18 new long-slit spectra of late-type galaxies from a large sample observed as part of a Tully-Fisher survey of galaxies in clusters with recessional velocities of $cz \sim 10000$ km s$^{-1}$ (Willick 1995). We refer to this sample as the aperture bias (AB) sample. These spectra were obtained with the Palomar 5-m telescope and double spectrograph. We use these data to quantify the effect of aperture bias on classifications made with small aperture spectra. The spectral coverage is from 3850 to 4700 Å and from 6450 to 7050 Å, with a resolution between 1 and 2 Å. The 1 arcsec wide slit was oriented along the major axis of the target galaxy. The exposure times are typically between 10 and 30 min. We have accompanying CCD images obtained at the Palomar 1.5m telescope from which we determine the morphological classifications and major axis orientation. These spectra are typically of lower S/N than the NG sample, but of higher S/N than the CL sample. To test how particular spectral features vary with galactocentric radius, we use the blue spectra to examine absorption line (Ca II H & K, and G-band) characteristics and the red to examine emission line (H$\alpha$) characteristics. These data are also not flux calibrated. An examination of aperture bias using this sample is presented in Appendix B.

The spectra in the CL and AB samples were reduced with IRAF in a standard manner for multifiber and long-slit data, respectively. The reductions consisted of bias subtraction, flat fielding, wavelength calibration, extraction, and sky subtraction (the last two steps are reversed for the long-slit data)[1].

## 3. CLASSIFYING GALAXIES

Our approach to spectral classification assumes only that galaxy spectra are a combination of (1) an old stellar spectrum, which we characterize as a K-star spectrum, (2) a young stellar spectrum, which we characterize as an A-star spectrum, and (3) an emission line spectrum. In the simplest terms, our classification procedure consists of measuring the relative contributions of these components to the observed galaxy spectrum and then comparing their relative strengths to those in the spectra of galaxies of known morphological type. We base this approach on the most striking spectral differences among the galaxies in the NG sample.

We use $\chi^2$-fitting to determine which linear combination of template spectra best reproduces the observed spectrum. Our set of templates consists of a composite[2] spectrum of a K star (characteristic of a stellar population $> 5$ Gyr old), a composite spectrum of an A star (characteristic of a stellar population $\lesssim$ few Gyr old), and an H II region (characteristic of ongoing star formation) divided into four separate template spectra: (1) the [O II]$\lambda\lambda 3726, 3729$ lines, (2) the H$\beta$ line, (3) the [O III]$\lambda\lambda 4959, 5007$ lines, and (4) the H$\alpha$ line. We excise from each composite stellar template the principal characteristic lines from the other template to construct two nearly "orthogonal" template spectra (the correlation between the final two stellar templates is 0.13). The modified K-star composite template includes all of the stellar absorption lines except for the hydrogen Balmer lines, while the composite A-star template includes only the Balmer lines. The H II region template is taken from an Irr galaxy (NGC 1569) in the Kennicutt sample. The quality of the CL data does not warrant the use of more detailed template spectra (*e.g.*, separate H & K and Mg$_2$ templates). We justify the choice of these templates over a mathematically orthogonal set derived from the data themselves (such as done by Connolly *et al.* ) by noting that for our low S/N spectra much of the variance is due to the noise, typically about one-half of the variance is accounted for by our best fit, so that a principal component analysis of the CL sample will not efficiently translate the mathematical properties of the sample into the physical properties of the underlying parameter space. A principal component analysis, such as that of Connolly *et al.* , on a large sample of high-S/N spectra would test the assumptions involved in our template selection.

We prepare each object and composite template spectrum before applying the $\chi^2$-fitting. First, we continuum subtract and standardize the template spectra to have a mean $\gg 0$ "counts" (for technical reasons) and a small maximum deviation from this mean (fixed, but arbitrarily chosen). The particular numbers in this procedure have no specific meaning, other than to insure the uniformity of the data. Second, the galaxy spectra are continuum subtracted and standardized to have an rms dispersion equal to the maximum deviation from the mean in the template spectra. Continuum subtraction, rather than division, is used because (1) continuum variations between objects over a limited wavelength region are primarily due to intrinsic brightness and color changes rather than detector sensitivity and (2) the low continuum levels in some objects would introduce significant noise. The spectra are rescaled to a fixed rms so that there is some standardization among template amplitude fits, but this is not a necessary step. The spectra are now ready for the $\chi^2$-fitter.

We fit our templates with the $\chi^2$-minimization fitting algorithm kindly provided by H.-W. Rix, which was developed by Rix & White (1992) to study detailed kinematics in elliptical and S0 galaxies. We fit the absorption line components first, by random walking through a parameter space that includes the recessional velocity, the Gaussian line-broadening function, and the relative contributions of the two absorption line templates. The best fit is then subtracted from the original spectrum. We then fit the emission lines to the residual spectrum, allowing the broadening function to differ from that used for the absorption lines. (The emission line broadening, which may arise from a few hot regions of narrow ($\sim 10$ km s$^{-1}$) linewidth, may differ from the stellar velocity dispersion (the terminology here includes bulk rotation) characterized by the absorption line broadening function.) The best-fit spectrum of a galaxy is then the sum of the absorption-line and emission-line fits created by minimizing $\chi^2$ on the two separate passes. Because of the two step fitting procedure, there is no possibility that emission and absorption lines would be constructed to cancel each other.

The $\chi^2$-fitting routine quantifies the contribution from each template, which we call template amplitudes, to the best-fit model spectrum for each galaxy. The template amplitudes are $\geq 0$. We calculate a goodness-of-fit criterion by dividing the rms deviation in the original spectrum by the rms deviation in the residual spectrum. The fit is acceptable if the rms residual is at most 70% of the original (a value determined from visual inspection to return reliable, S/N $\gtrsim$ 3, spectral features). Internal errors can be estimated from the range of models with a value of $\chi^2$ that is acceptable at some confidence level. One could propagate this uncertainty through the classification procedure to determine an internal estimate of the classification uncertainty. However, this is a complex procedure that relies on several estimated uncertainties and so we prefer to rely upon the external uncertainty estimates that are discussed further below. Figure 2 shows the fit for a random Sb galaxy in the Kennicutt sample (NGC 3147), whose rms residual is 65% of the rms in the original spectrum.

Several important aspects of the fitting routine are apparent in Figure 2. First, the continuum subtraction removes all structure on scales $> 200$ Å and in doing so does not alter any spectral line feature. Second, the model spectrum has almost no structure for $\lambda > 5500$ Å because our absorption line templates do not extend beyond 5500 Å. Third, the strengths of the absorption lines are somewhat underestimated. The lower right panel shows the fit to the Ca II H and K lines in greater detail. The slight discrepancy between the data and the fit is in part the result of the inversion of the ratio of H to K strengths between the K-star template and the object and of the imperfect matching of the width of the features. The latter may be the result of a non-Gaussian line profile, as one might expect from disk rotation in an Sb galaxy. Lastly, we do not include an [N II] template, so the fit around H$\alpha$ is often distorted in an attempt to account for some of the [N II] flux with the H$\alpha$ line. Despite these small flaws in the fit, we classify the galaxy as an Sb (as described in detail below), which matches its morphological classification (Kennicutt 1992).

---

[1] IRAF is distributed by the National Optical Astronomy Observatories, which are operated by the Association of Universities for Research in Astronomy, Inc. (AURA) under cooperative agreement with the National Science Foundation.

[2] The composite absorption-line templates are made from serendipitous observations of K-like and A-like stars with the Las Campanas multifiber system (S. Shectman 1995, private communication).

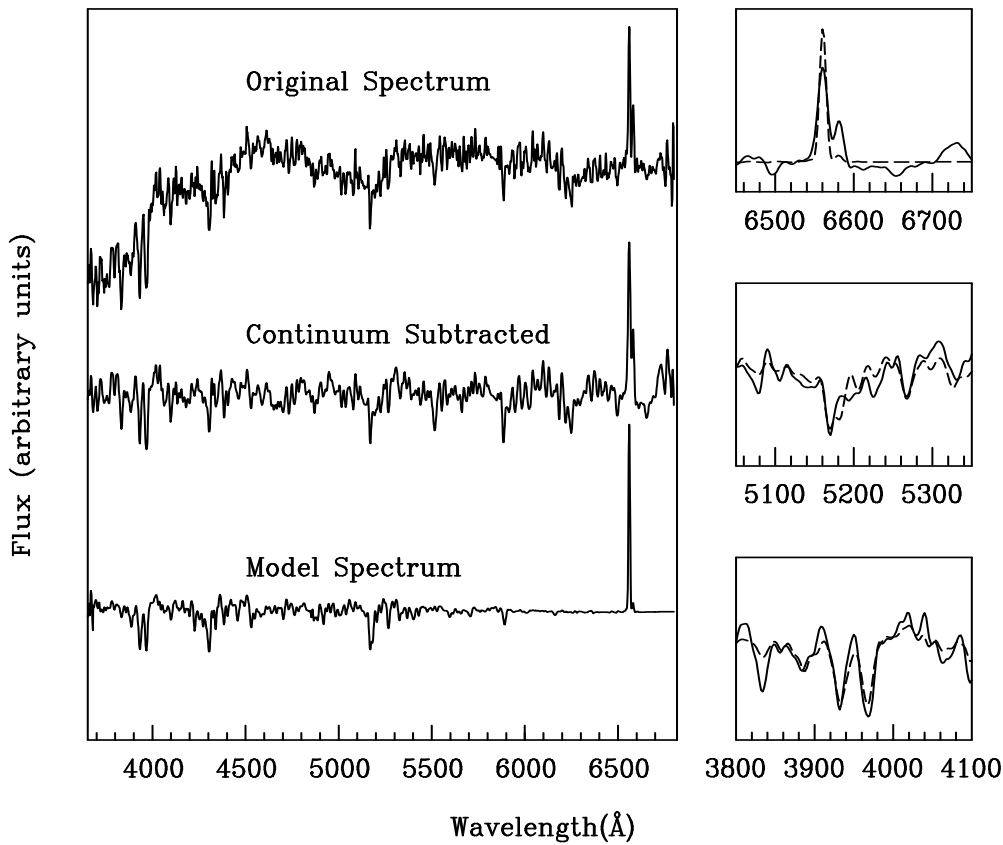

Figure 2: The original, continuum subtracted, and model spectra of a representative Sb galaxy (NGC 3147) from the NG sample. In the right panels, the dashed lines trace the fitted spectra.

Our spectral classification is based on three ratios: (1) the sum of the emission-line template amplitudes to the sum of the absorption-line template amplitudes, (2) the A-star template amplitude to the K-star template amplitude, and (3) the sum of the oxygen emission-line template amplitudes to the sum of the hydrogen emission-line template amplitudes.

These ratios were chosen because they isolate the most noticeable differences among the galaxy spectra in Figure 1 and represent physical components (ongoing star-formation, recent star formation, the dominance of an older stellar population, and the excitation of the gas) in the spectra.

The logarithms of the three ratios form a 3-dimensional space in which the galaxies are distributed. Our basic assumption is that galaxies of each morphological class inhabit a reasonably distinct region of this space — an assumption that is supported by the range of spectral characteristics seen in Figure 1 and by the results of Morgan & Mayall (1957). We define that region for a particular class of galaxy by calculating the positional centroid and the dispersion along each of the three coordinate axes for galaxies of that morphological type. Regions are defined for E, S0, Sa, Sb, Sc, and Irr galaxies using the published morphological classifications. A galaxy is spectrally classified by assigning it to the nearest group in this ratio space. The distance to any particular group is calculated by taking the distance along each of the three coordinate axis between the galaxy and the centroid of the group, dividing those distances by the dispersion of the group along that axis, and then adding the three components in quadrature. The distance along a particular axis is not used if that ratio is ill-defined; for example, if a galaxy does not have emission lines, then the ratio of oxygen to hydrogen emission is not used and that galaxy is classified within the 2-D space defined by the other two ratios. A projection of these regions is provided in §3.2 with the discussion of the classification of the CL sample.

Several technical details require further clarification. First, we use the NG and CL samples to self-calibrate the classification regions. Second, we use template amplitudes rather than equivalent widths because the continuum, which is often vary weak, would introduce large uncertainties into equivalent widths. Third, we use the logarithm of the template amplitude ratios to linearize differences among ratios. Fourth, when a template amplitude is less than 0.001, the template contribution is set at 0.001. This limit avoids divergent logarithms. Fifth, we determine whether a spectrum has emission lines by checking whether the largest emission line template amplitude is at least 0.20 (which translates typically into a $S/N \gtrsim 3$ in the peak of the line) or whether the median emission line template amplitude is greater than 0.001 (which implies that the program found two or more candidate emission lines corresponding to the absorption line redshift). Finally, we examined a variety of different combinations of template amplitudes ratios before deciding on the choices outlined above.

### a) THE NEARBY GALAXY (NG) SAMPLE

The NG sample, the simplest of the samples to analyze, provides the cleanest test of the method. After applying our spectral classification algorithm, 26 out of the 32 galaxies are classified to within one morphological type (81%). "Misclassifications" can arise from (1) erroneous morphological classifications, (2) errors in the fitting procedure, (3) errors in the classification procedure, and (4) scatter in the relationship between the morphological and spectral properties of galaxies. We neglect the first because it is beyond our control. The other sources of "misclassification" are the focus of the remainder of this paper.

There are several possible sources of errors in the fitting procedure. First, the templates have a slightly different spectral resolution than the NG data. Second, for simplicity we have not included a [N II] template, so the H$\alpha$ fits are sometimes distorted in an attempt to account for some of the [N II] emission. Third, the velocity broadening function is taken to be a Gaussian, which may lead to errors in the fit of galaxies with significant rotation. Fourth, we make no allowance for differences between the velocity widths of permitted and forbidden emission lines. Finally, we assume implicitly that all the absorption lines characteristic of the old population (e.g., Ca II H & K, $Mg_2$, Fe) are characterized by a single template spectrum. For all of these reasons, there is likely to be some template mismatch. We have chosen to minimize the number of templates and fitting functions in order to retain the stability of the procedure, rather than attempt to address some of these possible sources of uncertainty. These issues are mostly irrelevant for the low-S/N data of the CL sample because other sources of uncertainty dominate (cf. §3.2 and Appendices).

Uncertainties in the classification procedure are introduced by the small size of the NG sample. There are typically only three to five galaxies in each type bin. Therefore, the delineation of the regions of ratio space occupied by each morphological type is highly uncertain, and one peculiar object easily skews the classifications. Finally, we note that the same data are used both to classify and test the procedure. This approach may artificially elevate the success rate in such a small sample. These problems are less severe for the CL sample because it is a factor of 10 larger.

After inspection of individual fits and classifications for the NG sample, we conclude that the disagreements between the spectral and morphological classifications do not generally reflect problems in the quality of the data or the algorithm, but instead reveal a break in the detailed correspondence between spectral and morphological properties. Some of these galaxies are known to be morphologically peculiar. These galaxies are included in the NG sample because we chose to select our "normal" sample based solely on spectroscopic criteria, which we could apply to any sample of galaxies. The NG sample includes three objects that Kennicutt classifies as peculiar, one object that RSA classifies as amorphous, and the companion to M 51. Two of our six "misclassifications" are from this peculiar subsample.

The remainder of the "misclassifications" in the NG sample arise from scatter in the spectral characteristics within each morphological type. The regions occupied by galaxies of different morphologies in the 3-D ratio space overlap. It is therefore unavoidable that galaxies from one type will spill over into the region occupied by another type. We conclude, based on our inspection of the observed and model spectra, that the dispersion in spectral properties is real and not an artifact of the fitting procedure.

### b) THE CLUSTER (CL) SAMPLE

We now calibrate and test the classification method with the larger CL sample. Because the CL spectra do not extend beyond 6200 Å, we do not use the H$\alpha$ emission line template. Otherwise, the templates are the same as those used for the NG sample.

We find that the spectral and morphological types are consistent for 253 of the 304 sample galaxies (83%). This fraction is almost identical to that found for the much smaller NG sample. The centroids and standard deviations of the three template amplitude ratios (as derived from the NG and CL samples) are listed in Table 2 for each morphological type, and a projection of the regions occupied by the galaxies is shown in Figure 3. The ratios differ between the NG and CL samples because of the inclusion of H$\alpha$ in the analysis of the NG sample.

The oxygen to hydrogen emission line ratio has no discriminatory power for this sample (the same classification "success" rate is achieved without this ratio). Nevertheless, we retain it in the classification algorithm because it does vary among the latest types in the NG sample (cf. Figure 1) and would probably increase the number of successful classifications if the CL sample had a larger fraction of late-type spirals. For comparison, using only the emission to absorption template amplitude ratio produces a "misclassification" fraction of 0.25 for the CL sample and using only the A-star to K-star ratio results in a "misclassification" fraction of 0.61.

Table 2 : Spectral Classification Ratios from NG and CL Sample Galaxies

| Type | log(Em/Ab) | log(K Star/A Star) | log(O Em/H Em) | Number of Galaxies |
|---|---|---|---|---|
| NG Sample | | | | |
| (without H$\alpha$) | | | | |
| E | $-2.52 \pm 0.10$ | $-2.08 \pm 1.07$ | — | 4 |
| S0 | $-2.52 \pm 0.10$ | $-0.75 \pm 0.28$ | — | 3 |
| Sa | $-1.85 \pm 1.51$ | $-0.92 \pm 0.65$ | $-0.07 \pm 0.10$ | 5 |
| Sb | $-0.49 \pm 1.21$ | $-0.69 \pm 0.88$ | $0.75 \pm 0.92$ | 5 |
| Sc | $1.13 \pm 0.69$ | $-0.59 \pm 0.68$ | $0.16 \pm 0.23$ | 9 |
| Irr | $0.82 \pm 1.79$ | $-0.44 \pm 0.78$ | $0.42 \pm 0.22$ | 6 |
| CL Sample | | | | |
| E | $-2.42 \pm 0.61$ | $-1.52 \pm 1.10$ | $0.24 \pm 0.23$ | 52 |
| S0 | $-2.38 \pm 0.82$ | $-1.11 \pm 0.85$ | $0.55 \pm 1.68$ | 151 |
| Sa | $-1.83 \pm 1.24$ | $-1.06 \pm 0.98$ | $0.32 \pm 1.02$ | 40 |
| Sb | $-1.20 \pm 1.53$ | $-0.92 \pm 1.05$ | $1.32 \pm 1.65$ | 30 |
| Sc | $-0.23 \pm 1.06$ | $-0.50 \pm 0.87$ | $0.31 \pm 0.92$ | 26 |
| Irr | $0.11 \pm 0.19$ | $-0.48 \pm 1.31$ | $0.36 \pm 0.27$ | 5 |

In Figure 4, we plot the fraction of galaxies whose spectral and morphological classifications agree to within one type as a function of morphological type. The Figure illustrates that the CL sample is heavily biased toward elliptical and lenticular galaxies. The agreement between the spectral and morphological types is significantly greater at either end of the type range than in the middle because early-type spirals exhibit the largest variations in spectral properties (as evident in Figure 3), especially when H$\alpha$, which is brighter and less obscured by dust than H$\beta$, is excluded from the analysis. These intermediate-type galaxies have also been the most difficult to spectrally classify in previous studies (Connolly et al. and Bershady) and are the most difficult to classify visually. The similarity in the success rates between the NG and CL samples suggests that neither low S/N nor aperture bias has seriously affected our classification of the CL galaxies.

We now test the extent of the effects of low S/N and aperture bias on the classification of the CL sample. In the left panel of Figure 5, we plot a histogram of the classification success as a function of the goodness-of-fit criterion discussed previously (larger numbers indicate a better fit). This criterion reflects both the quality of the original spectrum (the S/N) and the success of the fitting algorithm, which in turn depends on such factors as template and velocity broadening mismatch. There is no evidence within the range spanned by the data that there is any dependence of the classification algorithm on this ratio. This result demonstrates that many "misclassifications" have not been introduced by the inclusion of poor quality spectra. The effect of noise on the fitting procedure is mostly to introduce spurious emission-line detections, which we minimized by the judicious selection of our emission-line criteria discussed at the beginning of §3. In Appendix A, we discuss adding noise to the NG spectra and demonstrate that the classification algorithm is reliable for S/N $\gtrsim 3$.

Because the galaxies in the CL sample are all observed with the same size aperture, the spectrum of a higher redshift galaxy includes light from a larger physical radius than does the spectrum of a lower redshift galaxy. To examine the effect of aperture bias on the CL sample, we plot a histogram of the classification success as a function of redshift in the right panel of Figure 5. The success rate is fairly constant across the range of redshifts and the misclassifications do not appear to depend on morphological type. Aperture bias does not systematically affect the classification of galaxies in the CL sample, and there appears to be no justification for excluding galaxies with $z < 0.1$ from this type of classification scheme. However, this statistical result does not prove that aperture bias is unimportant in the classification of any individual galaxy; in particular, we find that aperture bias can significantly distort some galaxy spectra (cf. Appendix B).

In Figure 6, we plot a comparison between the morphological and spectral classifications for the 51 galaxies that were "misclassified". Note that there are galaxies that are widely discrepant; for example, two morphological E's are spectrally classified as Irr's (see Figure 7 for the spectra of the three galaxies with the most discrepant spectral and morphological types). In future applications, it might be possible prevent the classification of a typical E as an Irr by introducing a magnitude criterion (or the velocity broadening determined with the $\chi^2$-fitter). There are a total of 13 morphological E's or S0's that are classified as spectral type Sb or later. The algorithm has "found" emission lines in all 13, although only eight have obvious emission lines (the remainder are marginal line detections). If the five galaxies without obvious emission are all spurious detections, the emission-line false detection rate is 2% (5 out of 203) among the E's and S0's. With our data, we cannot determine whether the eight galaxies with obvious emission are morphologically misclassified as early-types or are in fact ellipticals with some ongoing star formation (none have broad emission lines). The remaining widely discrepant galaxies are 12 late type galaxies (Sb or later) with strong H and K lines and no emission lines that we spectrally classify as either E's or S0's. It is unclear whether the spectral classification of these spirals is affected by aperture bias or reveals old disks with very low star formation rates.

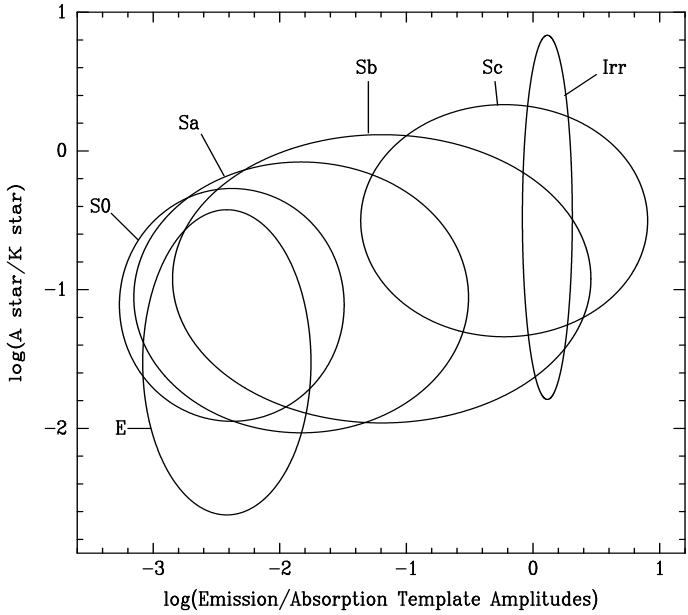

Figure 3: The spectral classification criteria derived from the CL sample projected onto the plane of emission/absorption and early/late template amplitudes. The ellipses are centered at the mean position for galaxies of each morphological type and are scaled to match the standard deviation along each axis of the galaxies within that type.

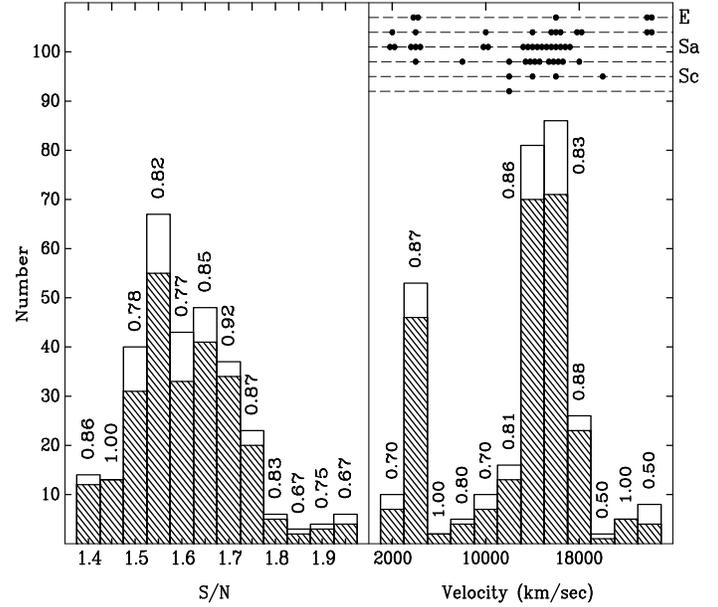

Figure 5: The classification success rates for the CL sample as a function of goodness-of-fit (as described in text) and redshift. The shaded regions represent the successful classifications. The rightmost bin in each panel includes all of the galaxies with characteristics that lie to the right of this bin. The fraction of successful classifications is given above each bin. In the upper portion of the right panel, we plot the morphological types of the galaxies misclassified in the corresponding bin.

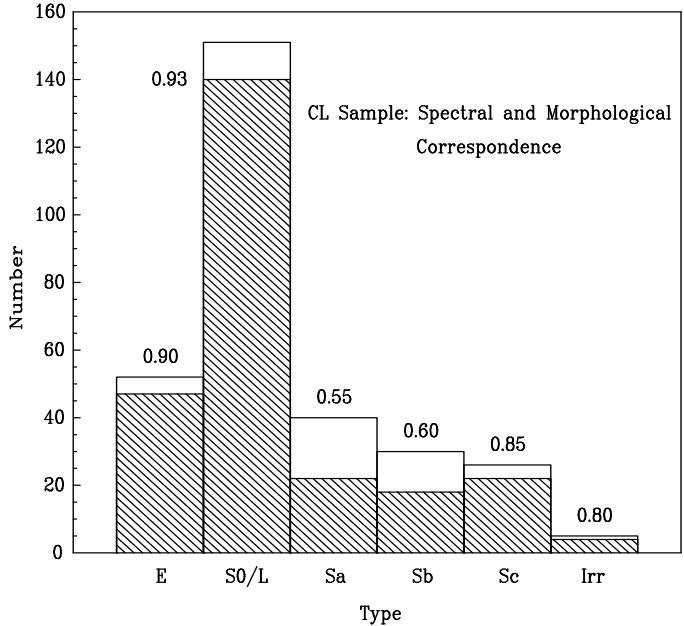

Figure 4: The fraction for which the morphological and spectral classifications agree to within one type as a function of morphological type for the CL sample. The shaded areas represent the successful fraction (the numerical fractions are also given).

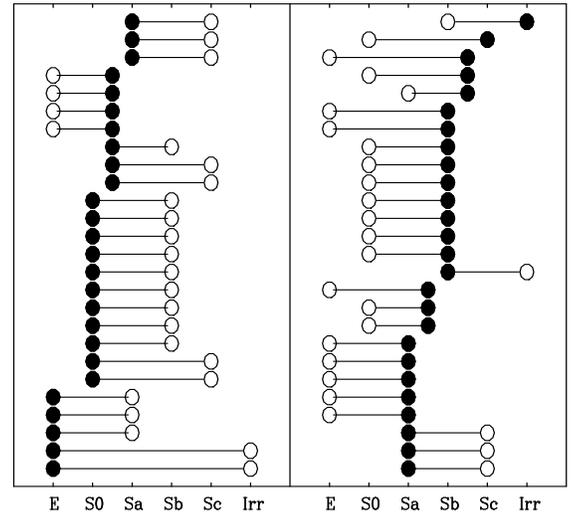

Figure 6: A comparison between the spectral versus morphological classification for the 51 galaxies from the CL sample for which the spectral type varied by more than one type from the morphological classification. The filled circles represent the morphological type from Dressler (1980) or Richter, Materne, & Hucthmeier (1987), and the open circles represent our spectral classification.

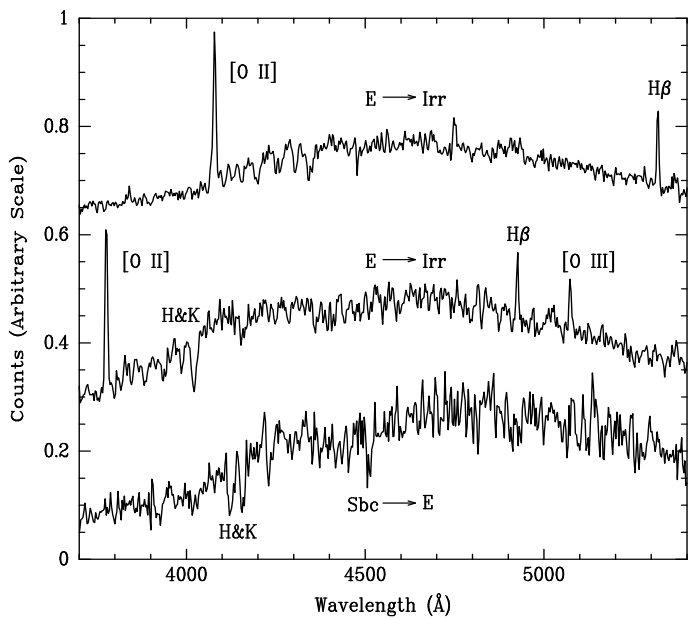

Figure 7: The spectra of the three galaxies in the CL sample with the most discrepant spectral and morphological types. Major spectral lines are labeled. The morphological type is shown to the left of the arrow and the spectral type to the right. The spectra have been smoothed by a factor of 4.

These results, and those for the NG sample, indicate that the spectral and morphological classifications disagree by more than one type for only about one-fifth of the galaxies. As mentioned in §3.1, this inconsistency can arise for a variety of reasons. The interesting scientific issue is to what degree these mismatches result from a break in the correlation between morphological and spectral characteristics. Resolving this issue requires a detailed knowledge of the contribution of the possible sources of error. Based on the arguments presented in the Appendices and the very similar misclassification fractions in the NG and CL samples, we conclude that most of the misclassifications arise from real scatter in the spectral properties of galaxies within each morphological bin. Some of that scatter is relatively small as evidenced by the overlap seen in Figure 3 (only 24% of the CL misclassifications are due to mismatches of more than 2 types). However, there is a small fraction of galaxies for which the spectral and morphological types are grossly different, such as the morphological E's with emission lines. Such objects require further study and demonstrate why spectral and morphological classifications are complementary.

Lastly, we present one more test of the classification algorithm. We use the classification calibration derived from the CL sample to classify the NG sample. We recalculate the various template amplitude ratios for the NG sample excluding H$\alpha$, since H$\alpha$ is not used in the CL sample. Applying the CL criteria we correctly classify 91% of the NG sample. This success rate is *higher* than that attained with self-calibration (cf. §3.1). This result partially addresses the concerns that the lack of flux calibration severely distorted the template amplitude ratios and that the success rate has been elevated by using self-calibration. We also present in Table 2 the calibration derived from the NG sample excluding H$\alpha$ to enable a direct comparison of the derived criteria from the two samples. The values from the NG sample have larger external uncertainties due to the small number of objects per bin. We conclude by stressing that the applicability of the classification criteria derived for the CL to other samples has not been fully investigated. Any application of this technique to other samples must be self-tested.

4. SUMMARY

We develop and test a method of spectroscopically classifying nearby galaxies. The strengths of the method are that (1) for each galaxy, we obtain an estimate of the relative contributions of the populations of young and old stars, (2) unlike morphological classification, the method is independent of spatial resolution, (3) the method can be applied uniformly to existing data for thousands of galaxies, (4) the classification uncertainties are quantifiable through Monte-Carlo simulations and comparison with morphological classifications, and (5) the method is independent of continuum shape and so does not require fluxed spectral data. There is a strong correspondence between spectral classification and visual morphological classification ($\gtrsim$ 80% of galaxies can be spectrally typed to within one morphological type). The spectral types might correspond even more strongly to other galaxy properties that are known to correlate with Hubble type (*e.g.*, HI content, environment, or luminosity), and an examination of those correlations can provide greater insight into galaxy evolution. In future applications, the independence of our method on continuum shape might also be utilized to test the method against the other classification methods that depend on continuum shape.

The principal applications of spectral classification algorithms are to high-redshift surveys, where the limited spatial resolution prohibits visual classification, and to large redshift surveys. Whether one uses colors, continuum shape, or spectral line properties as the basis of spectral classification depends primarily on the type of data that is available. It is now possible to determine the local density-type, luminosity-type, and correlation function-type relations from large redshift surveys without visually classifying tens or even hundreds of thousands of galaxies. In addition, these same relationships can be examined in terms of spectral characteristics, such as the presence of a young stellar population.

We define and test our algorithm with three samples. The first is a small, but otherwise ideal sample (NG) that does not suffer from aperture bias and has high S/N. The results are used to illustrate the success of the basic classification procedure. The second sample (CL) consists of low S/N fiber spectra of galaxies of known morphological type taken from a redshift survey with a multifiber spectrograph. These spectra are of similar quality to those obtained from large redshift surveys (e.g., Huchra et al. 1992; Shectman et al. 1992). We find that despite an important limitation of the data, the lack of spectral coverage at the H$\alpha$ line, our spectral classification matches the morphological type for $\gtrsim$ 80% of the CL galaxies (the same as the fraction "correctly" classified in the NG sample and consistent with the results of Morgan & Mayall 1957). With these two samples and a third sample (AB) of long-slit spectra, we quantify the effects of low S/N and of aperture bias on the classifications. We find that the algorithm performs well in the low S/N regime (S/N $\gtrsim$ 3) and that aperture bias does not seriously affect the fiber data. An amalgamation of continuum and line spectral classification methods should strengthen the classification procedure (although fluxed spectra are usually not available from large redshift surveys). Finally, we stress that although the classification method is general, the amplitude ratios and the particular classification criteria we derive are not, because the CL data are not flux calibrated.

The correspondence between spectral and morphological classifications places a strong constraint on models of galaxy formation and evolution, because the absence of large populations of very young E's or very old S's is still unexplained. The well-known decrease in star formation activity along the Hubble sequence is clearly visible in the second and third columns of Table 2. In the progression from early- to late-type galaxies, both the mean emission to absorption line ratio and the A star to K star ratio systematically increase (although the scatter is large).

The comparison between the spectral and morphological classifications also reveals some unusual galaxies. For example, the samples include ellipticals with line emission and late-type spirals in which star formation appears to have ceased. In total, about 10% of E's and possibly a similar number of spirals (depending on the degree of aperture bias) have more in common with galaxies of the opposite spectral type than with their own. Whether these unusual objects result from recent galaxy-galaxy interactions or represent the respective tails of the evolutionary histories of elliptical and spiral galaxies is unknown. This issue may be resolved soon, when the use of automated spectral and morphological classification (Abraham et al. 1994; Han 1995) makes possible a detailed study of the properties and environments of thousands of galaxies.

Acknowledgments: The authors wish to thank M. Bershady for a detailed reading and comments, R. Kennicutt for making available his compilation of galaxy spectra and for insightful comments on a preliminary draft, and H.-W. Rix for stressing the importance of aperture bias and providing his $\chi^2$-fitting software. DZ acknowledges partial financial support from NASA through HF-1027.01-91A from STScI, which is operated by AURA, Inc., under NASA contract NAS 5-26555. AIZ acknowledges support from the Carnegie and Dudley Observatories.

APPENDIX A: SPECTRAL CLASSIFICATION IN THE LOW S/N REGIME

Most of the data to which our spectral classification method might be applied are low ($\lesssim 10$) S/N spectra obtained originally for redshift measurements. We now test the limit in S/N for which we can obtain reliable classifications by degrading the NG spectra to even lower signal-to-noise than is characteristic of the CL sample.

We create a series of noise-enhanced versions of the NG spectra by adding Gaussian uncorrelated noise to each pixel and degrading the typical S/N to 3.5, 2.3, and 1.8 in three different tests. These S/N values are based on the ratio of the *depth* of the Ca II absorption lines to the noise — not the ratio of the integrated signal in the features to the noise. We prefer to use the feature depth rather than its equivalent width because many of the spectra have low continuum levels which introduce large uncertainties into the equivalent width. The fitting and classification are then redone by using the regions in ratio space defined by the original NG data. Decreasing the S/N to 2.3 is sufficient to boost the fraction of spectra for which the spectral and morphological classifications disagree by more than one type to above 30%, which roughly corresponds to a $1\sigma$ uncertainty per type in

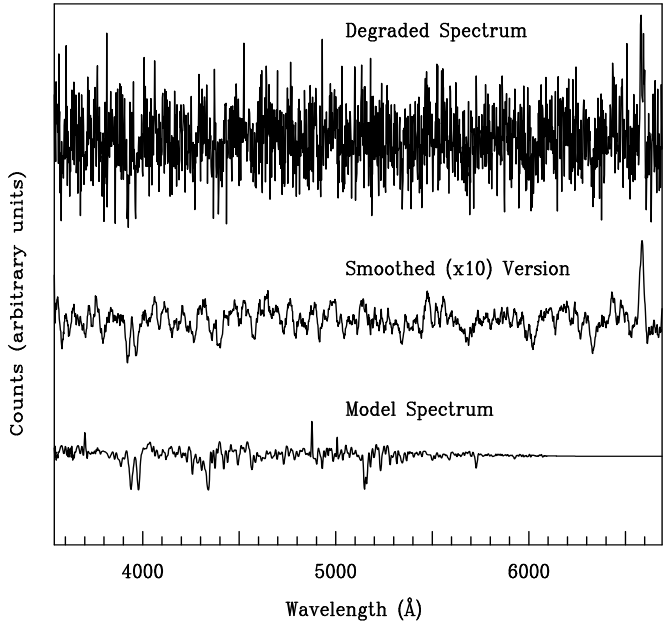

Figure 8: The degraded (S/N = 2.3), smoothed, and model spectra for NGC 3147. This plot represents the level of noise at which significant (> 30%) misclassification occurs. The spurious emission lines that produce the misclassification can be seen in the model spectrum.

We quantify how the spectrum changes with noise by calculating the cross-correlation of the original and noise-enhanced spectra. The mean correlation value, for the S/N = 2.3 case, is 0.59, and so the noise has significantly altered the spectrum. Figure 8 shows the noise-enhanced version of the spectrum of NGC 3147 shown in Figure 1. We also plot a smoothed version to demonstrate that there is still significant signal in the spectrum, and then plot the model fit. In this case, the problem is that noise spikes are easily identified as [O II], H$\beta$, and [O III] emission lines (the H$\alpha$ emission line is not fit because the wavelength range for the CL sample does not extend that far redward). We conclude that S/N $\gtrsim 3$ (peak to noise) is necessary to maintain the standard 80% correspondence between morphological and spectral properties (cf. Morgan & Mayall 1957; our NG and CL results).

APPENDIX B: APERTURE BIAS

Aperture bias is a potentially serious problem for spectral classification based on multi-object observations. Fiber apertures are typically about 2 arcsec (although the LCO spectrograph has 3 arcsec apertures), and multislits are usually also small. At recessional velocities as large as 30000 km s$^{-1}$ (z = 0.1), a standard 2 arcsec aperture samples a radius of only 2 kpc.

Our investigation of aperture bias focuses on the AB sample. This sample includes only spirals, which are generally more susceptible to aperture bias than ellipticals, because spiral spectral properties are typically less homogeneous as a function of radius (*e.g.*, bulge, disk, HII regions, and dust). With long-slit data, we can vary the effective aperture and thus the area of the disk that we use to classify the galaxy.

The AB sample consists of 18 galaxies from two galaxy clusters, Abell 2199 (z = 0.03) and Abell 2657 (z = 0.04). For each galaxy, there is a blue and red spectrum (cf. §2). With these spectra, we quantify the variations in the emission and absorption lines as a function of radius. Because of the restricted wavelength range of these data, especially the absence of the region from 4700Å to 5100Å, we cannot apply the classification algorithm directly. Instead, we measure the change in each spectrum as the effective aperture is varied. Each spectrum is reduced in a standard manner through the sky subtraction step. We then extract different width apertures that trace the galaxy spectrum. The aperture width is incremented by 2 pixels for each extraction until only noise is being added to the total spectrum (as defined by the first net decrease in the total counts in the spectrum). For each galaxy, we cross-correlate each of the extracted spectra with the widest-aperture spectrum to estimate the difference between the two.

Ideally, we would have spectra that fully sample the galaxy, rather than long slit spectra. For face-on galaxies, one could multiply the flux observed at radius $R$ within the slit by $R$ to get the flux in an annulus. However, the AB galaxies (by virtue of being candidates from a Tully-Fisher study) are at large inclinations to the line-of-sight. Any transformation from the flux within the slit to total flux requires not only azimuthal corrections, but inclination corrections requiring assumptions about the optical depth of the disk.

Such corrections to the noisy spectral data at large radius are dubious. Therefore, we make no correction to the data as a function of radius, but note that the contributions to the flux from large radii may be underrepresented.

Differences among the extracted spectra of a galaxy can arise from changes in the S/N. To account for a decrease in the cross-correlation value due to noise, we first construct a simulated narrow aperture spectrum by adopting the reference (wide aperture) spectrum as the true spectrum within the narrow aperture (*i.e.*, we assume that the signal does not change across the entire aperture and that the reference spectrum is noiseless). To correct the signal strength for the change in aperture size, we divide the simulated spectrum by the ratio of the total counts in the reference to those in the observed narrow aperture spectrum. We subtract a simulated sky (with a mean of 0 counts and a noise level given by the readnoise $\times\sqrt{\text{width of the aperture in pixels}}$) and add Poisson noise to the signal. The cross-correlation between the simulated spectrum ($S$) and the reference spectrum ($R$) then quantifies the degradation of the cross-correlation due to a change in the S/N. The cross-correlation coefficient for the observed narrow aperture spectrum ($N$) and the reference is standardized using the previous estimate of the degradation due solely to noise. The corrected correlation values are

$$C = \frac{\sum_i N_i R_i \sum_i S_i S_i}{\sum_i S_i R_i \sum_i N_i N_i},$$

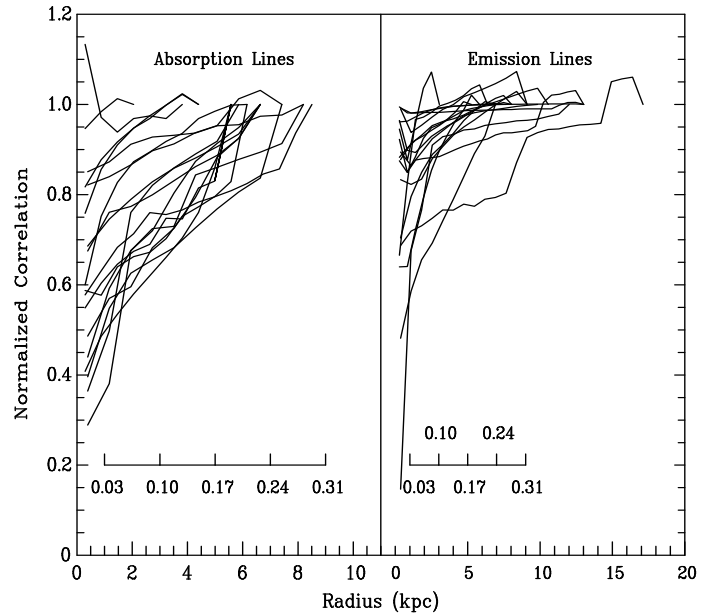

Figure 9: The corrected correlation as a function of aperture size for the AB sample (definition of corrected correlation is given in Appendix B). The scales within the plots show the corresponding physical size of a 3 arcsec aperture for a galaxy at the labeled redshift.

where the sum over $i$ represents the sum over individual pixel values and each spectrum has had a continuum fit subtracted. The values of $C$ for the absorption line spectra as a function of aperture size are shown in the left panel of Figure 9. Also displayed is a guide that indicates the physical size probed by a 3 arcsec aperture at various redshifts.

As Figure 9 shows, the cross-correlation values for the absorption line spectra (the blue spectra) drop for most galaxies as the aperture narrows. This result implies, not surprisingly, that the spectra of the inner regions of some galaxies are different than the integrated spectra. The decrease in the correlation value at its most extreme is about 50%. In Appendix A, we showed that lowering the mean correlation value to 59% (S/N = 2.3) reduced the fraction of successful classifications to below 70%. For redshifts $z < 0.03$ (9000 km s$^{-1}$), the Figure implies that aperture bias *may* significantly distort the classifications of some spiral galaxies. However, there are no systematic errors in the classifications of the CL sample at low redshifts. This success results partly from the domination of the sample by early-type galaxies and partly from the classification of a galaxy as late-type regardless of its absorption spectrum if it has strong emission lines.

Several other factors affect the cross-correlation values. For example, there are differences in the line profiles between the inner and integrated apertures. Line profile differences reflect kinematic changes and have a smaller effect on the classification than differences in line intensity ratios. For example, in cases where rotation causes a bi-modal line profile, the classification will be unaffected if both the early-type and late-type components are similarly distorted, but the cross-correlation value will be affected. Imperfect sky subtraction may also contribute to the decrease in the correlation value in at least several cases. Therefore, it is an overestimate to attribute all of the decline in the cross-correlation values in Figure 8 to differences in the radial distribution of stellar populations.

We treat the emission line spectra (the red spectra) slightly differently. We only consider the region of each spectrum around the H$\alpha$ and [N II] lines. The rest of the procedure is identical to that used for the absorption lines. The results are shown in the right panel of Figure 9. Most of the galaxy spectra change little between narrow and wide apertures, which indicates that the galaxies in this sample have emission lines (or lack them) throughout their disks.

One notable exception is not well illustrated by the Figure. In this galaxy, the continuum emission drops below sky at a radius at which no line emission is detected, and so the galaxy appears to have similar spectral properties at all radii. However, the 2-D spectrum reveals strong H$\alpha$ emission well beyond the radius where the continuum emission is undetectable. This result is a clear example of aperture bias. The bias against observing this line emission would occur even if the galaxy had a redshift of several tenths. For such a galaxy, the emission lines that would identify it as a spiral will not be found in either multifiber or multislit data. Although there will always be some fraction of spirals that are misclassified due to this problem, twelve of the eighteen spirals we examine have easily detectable emission lines, and, of those, only the one has emission exclusively at large radius.

We conclude from Figure 9 that the spectral classification of spiral galaxies without detected emission lines and with recessional velocities below 15000 km s$^{-1}$ has about a 50% chance of being significantly affected by aperture bias. Combining this percentage with the observation that emission lines (affected by aperture bias in one out of 12 galaxies) are detected in 70% of the CL sample spirals of morphological type Sb or later, we expect that aperture bias may be a problem in about 20% of spirals with $z \lesssim 0.05$. Ellipticals and lenticulars, which are much more radially homogeneous, are even less likely to be affected. Aperture bias does not appear to be a significant problem in the CL survey, although it may explain the 12 morphologically late-type galaxies that are spectrally classified as E's or S0's.

We conclude that, for 3 arcsec apertures, aperture bias does not seriously affect classification for the vast majority of galaxies, even those at redshifts $z \sim 0.03$. These results are general trends, and we stress that aperture bias may still be very important in certain cases.